\documentclass{emulateapj}

\shorttitle{Missing Supergiants in the LMC}
\shortauthors{DESGW Team}

\usepackage{textcomp}
\usepackage{changes}

\begin{document}
\title{A Dark Energy Camera Search for Missing Supergiants in the LMC
  After the Advanced LIGO Gravitational Wave Event GW150914}


\author{
J.~Annis\altaffilmark{1},
M.~Soares-Santos\altaffilmark{1},
E.~Berger\altaffilmark{2},
D.~Brout\altaffilmark{3},
H.~Chen\altaffilmark{4},
R.~Chornock\altaffilmark{5},
P.~S.~Cowperthwaite\altaffilmark{2},
H.~T.~Diehl\altaffilmark{1},
Z.~Doctor\altaffilmark{4},
A.~Drlica-Wagner\altaffilmark{1}
% ...
\\ \vspace{0.2cm} (The DES Collaboration) \\
}
 
\altaffiltext{1}{Fermi National Accelerator Laboratory, P. O. Box 500, Batavia, IL 60510, USA}
\altaffiltext{2}{Harvard-Smithsonian Center for Astrophysics, 60 Garden Street, Cambridge, MA, 02138}
\altaffiltext{3}{Department of Physics and Astronomy, University of Pennsylvania, Philadelphia, PA 19104, USA}
\altaffiltext{4}{Kavli Institute for Cosmological Physics, University of Chicago, Chicago, IL 60637, USA}
\altaffiltext{5}{Astrophysical Institute, Department of Physics and Astronomy, 251B Clippinger Lab, Ohio University, Athens, OH 45701, USA}
% ...

\begin{abstract}
  The collapse of the core of a star is expected to produce gravitational
  radiation.  While this process will usually produce a
  luminous supernova, the optical signatue could be subluminous and 
  a direct collapse to a black hole, with the star just disappearing, is possible.
  The gravitational wave event GW150914 reported by the
  LIGO Virgo Collaboration  (LVC) on 2015 September 16, was detected by a
  burst analysis 
  and whose high probability spatial localization
  included the Large Magellanic Cloud.  
  Shortly after the announcement of the event, 
  we used the Dark Energy Camera to observe 102 deg$^2$ of the
  localization area, including a 38 deg$^2$ area centered on the LMC.  
  Using a catalog of 152 LMC luminous red supergiants, candidates to
  undergo a core collapse without a visible supernova, we 
  find that the positions of 144 of these are inside our images,
  and that all are detected --- none have disappeared.
  There are other classes of candidates: we searched existing catalogs of
  red supergiants, yellow supergiants, Wolf-Rayet stars, and luminous blue
  variable stars, recovering all that were inside the imaging area.
  Based on our observations, we conclude that it is unlikely that 
  GW150914 was caused by the core collapse of a supergiant in the LMC, 
  consistent with the LIGO Collaboration analyses of the gravitational waveform
  as best described by a high mass binary black hole merger.
  We discuss how to generalize this search for future 
  very nearby core collapse candidates.
\end{abstract}

\section{Introduction}

On 2015 September 14 the Advanced LIGO interferometer network detected
a high significance candidate gravitational wave (GW) event
(designated GW150914; \citealt{ligo2016}) and two days later provided
spatial location information in the form of probability sky maps
\citep{gcn1}.  The analysis that produced the trigger was sensitive to
bursts, suggested a high source mass, and yielded localization contours that
enclosed the Large Magellanic Cloud (LMC) at high confidence.
Burst-like gravitational wave signals could originate from the
core-collapse of massive stars, and there is evidence that $\sim20\%$
of core-collapse events fail to produce a luminous supernova (SN); see
for example, \citep{kochanek2015}.

Motivated thus, we initiated observations of the LMC with DECam on
2015 September 18 in an effort to search for a potential failed SN
through the disappearance of a massive star.  We select 152 high
luminosity supergiants that are candidates for becoming failed
supernova, locate and verify that the 144 inside our DECam data are
still present after the LIGO event, making it unlikely that GW150914
originated from a failed SN in the LMC.  In January 2016 an improved
analysis of the LIGO data for GW150914 changed both the spatial
localization (moving it away from the LMC) and the source model 
(now shown to be consistent with a binary black hole merger by \citet{ligo2016});
this GW source did not originate from the death of a massive star in the LMC.
Our analysis, however, represents an important template for the follow up of future 
burst-like GW events coincident with very nearby galaxies.
 
\section{LIGO event GW150914}

On 2015 September 14 at 09:50:45 UT the Advanced LIGO interferometers
at Hanford and Livingston recorded burst candidate event GW150914
during Engineering Run 8.  This event was triggered by the cWB
(coherent WaveBurst) unmodeled burst analysis during real-time data
processing.  On 2015 September 16, the LIGO Virgo Collaboration (LVC)
provided two all-sky localization probability maps for the event,
generated from the cWB and LALInferenceBurst (LIB) analyses
\citep{gcn1}.  The cWB online trigger analysis makes minimal
assumptions about signal shape by searching for coherent power across
the LIGO network \citep{klimenko2008}.  The LIB analysis is a version
of the the LALInference analysis \citep{veitch2015} Bayesian
forward-modeling-based follow up tool that uses a Sine-Gaussian signal
morphology instead of models of compact binary mergers; for
information on both algorithms see \citet{essick2015}.  No
LALInference detection using a compact binary mergers model was
announced.  Stellar core collapses cause significant signals in the
cWB analysis (but not in LALInference) though the core collapse would
have to be nearby \citep{fryer2011, gossan2015}.

The LVC released localization sky maps of the GW150914 event to
make possible electromagnetic follow-up of the GW150914 event (\citealt{ligoEM};
see also \citealt{aasi2014}).
The maps provided
spatial localizations of 50\% and 90\% confidence regions encompassing
about 200 and 750 deg$^2$, respectively.  The area enclosing 50\% of
the total probability passed through the center of the Large
Magellanic Cloud, a $0.2$ L$^{\star}$ galaxy at a distance of 50 kpc
\citep{walker2012, degrijs2014}: see the dotted lines showing the
enclosed cWB sky map probability in Figure~1.  The high probability
ridge line passed over 30 Doradus and the proto-globular cluster R136.

We recently began an observational program
using the wide-field Dark Energy Camera (DECam; \citealt{flaugher2015}) on the
Blanco 4-m telescope at Cerro Tololo Inter-American Observatory to
search for optical counterparts to GW triggers. Our wide-field 
search for counterparts to GW150914 is described in the companion paper
\citet{marcelle2016}; an overview of the program is in \citet{des2016}.  
We additionally designed a specific set of
observations to search for failed SNe in the LMC, using 5-sec $i$ and
$z$ band observations covering 38 deg$^2$ centered on the LMC on 2015
September 18 and 27, in seeing of 1.1--1.3$\arcsec$.  

Subsequently, on 2015 October 3, the LVC revised its analysis: the
data were most consistent with a binary black hole merger \citep{gcn2}.
On 2016 January 13, the LVC provided new skymaps, the
most accurate and authoritative of which was the LALInference analysis
\citep{gcn3}.  The new contour enclosing 50\% of the total probability
shifted southward of the LMC, although the LMC is still inside the
90\% contour.

\begin{figure}
\hspace*{-0.5cm}\includegraphics[width=0.52\textwidth,trim=0cm 0 0 0]{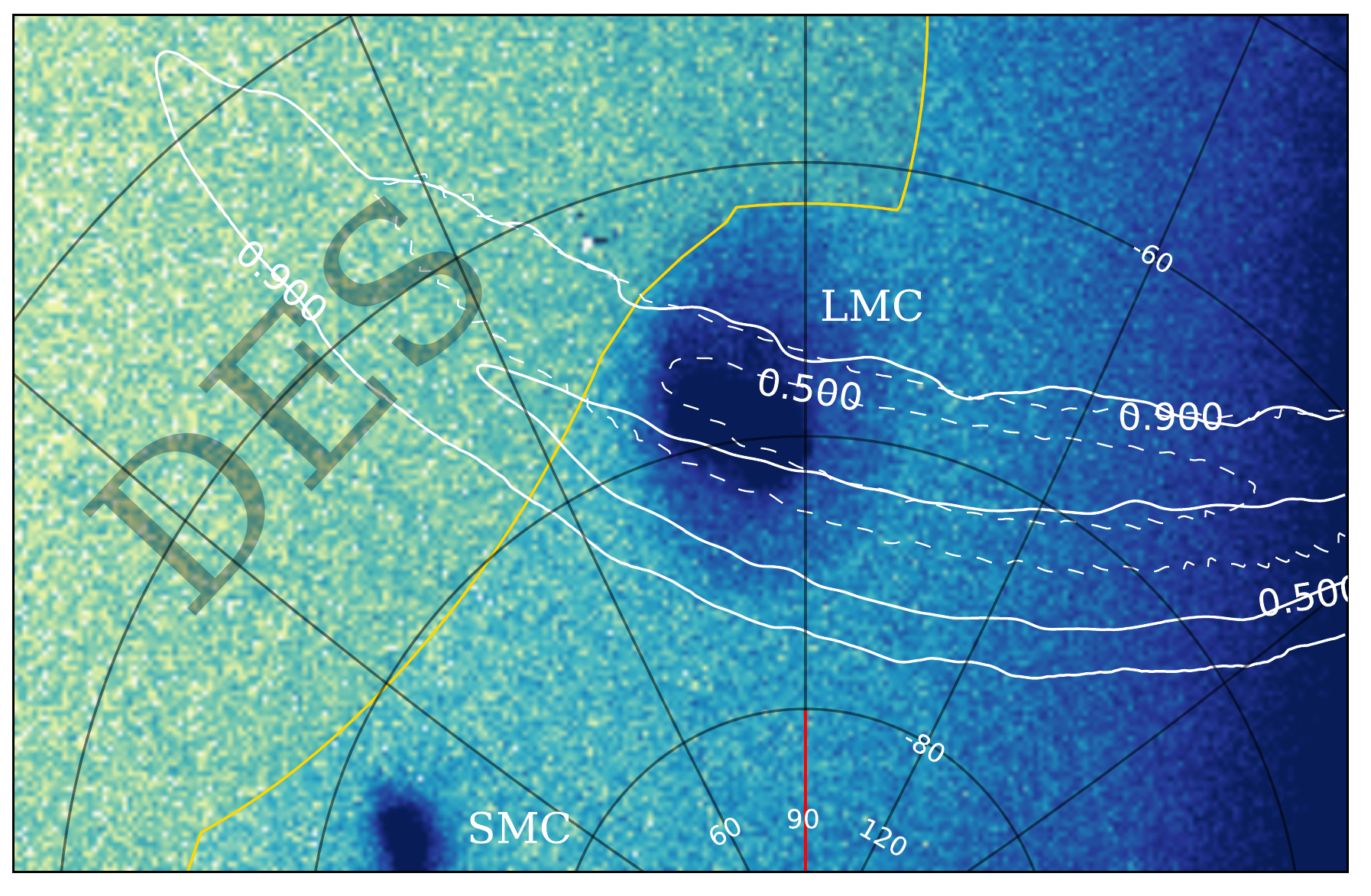}
\label{fig:contours}
\caption{A map of the logarithm of 2MASS J-band star counts
   around the LMC with the
  LIGO localization contours shown in white.  The contour labels
  indicate the fraction of the LIGO localization probability enclosed.
  The dotted contours are for the initial (Sept 2015) {\tt
    skyprobcc\_cWB\_complete} map, while the solid contours are for
  the final (Jan 2016) {\tt LALInference\_skymap}.  There is an island
  of significant probability in the Northern hemisphere in the {\tt
    skyprobcc\_cWB\_complete}, not present in the {\tt
    LALInference\_skymap}, so the dotted contours do not show the
  complete 50\% or 90\% areas. The data
      are shown on an equal-area McBryde-Thomas flat-polar quartic
          projection, as is Figure 3. }
\end{figure}

\section{Core-Collapse Signatures}

A normal core-collapse SN in the LMC is a remarkably obvious event---
SN1987A was found by eye as a new 5\textsuperscript{th} magnitude
object 24 hours after the core collapse.  Core-collapse SNe have peak
absolute magnitudes of $\sim$ -21 to $\sim$ -14, which at the distance of
the LMC corresponds to apparent magnitudes of -2.5 to 4.5.  

However, it has been argued that up to $\sim$20\% of core-collapse SNe
are not optically luminous \citep{kochanek2008}, and there is
recent evidence that luminous supergiants specifically
are prone to be failed SNe.  Two candidates are
currently known: the Large Binocular Telescope survey
\citep{gerke2015} found a $18-25$ M$_\odot$ star missing, and a Hubble
Space Telescope archival survey \citep{reynolds2015}, found a $25-30$
M$_\odot$ star missing.  These objects are sufficiently nearby
that a SN associated with the event would have been detected, by the Large
Binocular Telescope survey itself in that case.  In addition, the
population of known progenitors to Type IIP SNe lacks red supergiants
above $\gtrsim 17$ M$_\odot$ \citep{smartt2009}, suggesting that
that more massive red supergiants end in a failed SN.  This line of
argument reproduces the current black hole mass function \citep{kochanek2015};
similarily the purely theoretical study of core collapses by \citet{sukhbold2015}
reproduces both the neutron star and black hole mass functions.
Pre-collapse, red supergiants are very luminous:
\citealt{smartt2015} shows that the missing SN progenitors have
 $\gtrsim 10\textsuperscript{5.1}$ L$_\odot$.

\section{Optical Signatures of a Failed Supernova}

There are three viable signatures for a failed supernova: (1) the star
might simply collapse to a black hole; (2) the unbound outer
atmosphere of the star may expand and cool, gaining in luminosity as
it expands; and (3) there might be a shock from the creation of the
neutrinosphere that propagates through the atmosphere to the outer
layer, causing a shock breakout flash.

We briefly discuss these potential signatures here.  The hydrogen
atmospheres of these supergiants are so marginally bound to the star
that the creation and free streaming of the neutrinosphere during
core-collapse may remove enough mass to unbind the atmosphere
\citep{nadezhin1980}.  If the shock from the neutrinosphere creation
is energetic enough it will cause the unbound atmosphere to expand,
necessarily cooling and gaining in luminosity as it expands.
\citet{lovegrove2013} simulated this process using realistic models of
15 and 25 M$_\odot$ red supergiants, finding that the transient is
long ($\sim$ year, 10\textsuperscript{3} K, 10\textsuperscript{6}
L$_\odot$), and that the unbinding of the atmosphere was more likely
in the 15 M$_\odot$ than in the 25 M$_\odot$ star.  
The shock breakout signature was studied by \citet{piro2013} who found that it 
would present a short, hot transient ($\sim$ week, 10$^4$K, 10$^{6.5}-10^{7.5}$ L$_\odot$).
At the distance of the LMC this would be remarkably bright: $i\approx$ 5.1 - 7.6
(see Table~\ref{table:sig}).  The existence of a shock breakout does, however,
depend on sufficient energy in the shock, and this is unclear.

The Nadezhin brightening of signature 2 lasts hundreds of
days, with a lower bound in luminosity of the pre-collapse luminosity
of the star, but possibly rising to $L \sim 10^{5.5}-10^{6.5}$
L$_\odot$, presumably with an effective temperature starting close to
the pre-collapse star and cooling thereafter.  At the distance of the
LMC, this is $i\sim$ 6.7 - 9.3.  These objects would look much like
the supergiant has brightened by a couple of magnitudes.  

\begin{deluxetable}{lcccrr}
\tablecolumns{5}
\tablecaption{Predicted optical signatures of 
a failed supernova in the LMC\label{table:sig}}
\tablehead{
    \colhead{} &
    \colhead{$i$} & \colhead{$(g-i)$} &
    \colhead{$K$} & \colhead{$(J-K)$} & \colhead{timescale}}
\startdata
supergiants & 8.0-11.5  & 1.5-2.3    & 6.0-8.0 & 0.9-1.4 & $\gg$ 1 year \\
shock break out\tablenotemark{a}  
    & $\sim$ 5.1-7.6 & $\sim$ 0.2 & $\sim$ 4.6-7.1 & $\sim$ 0.07 & $\sim$ 1 week\\
Nadezhin\tablenotemark{b}    
    & $\sim$ 6.7-9.3 & $\gtrsim$ 1.5 & $\sim$ 4.6-7.1 & $\gtrsim$ 0.9 & $\sim$ 1 year\\
disappear   & --- & --- & ---       & ---  \\
\enddata
\tablenotetext{a}{Assuming a blackbody spectrum}
\tablenotetext{b}{Assuming a supergiant-like spectrum}
\end{deluxetable}

\section{LMC Red Supergiants}

Our search  focuses on high luminosity red supergiants in the
LMC; we will consider other candidate failed supernova
progenitors in the next section.  
The two best studies of large numbers of LMC supergiants are by
\citet{neugent2012} and \citet{gonzalez2015}.  Both combine 2MASS
point source data \citep{2mass2006} with astrometric catalogs (UCAC-3
or USNO-B1; \citealt{monet2003}), using proper motions to reject Milky
Way (MW) stars, and then using infrared colors and $K$
magnitudes to select the supergiants.  Both studies performed spectroscopy for their
final identifications.\footnote[1]{We will drop the proper subscript $s$ from the 2MASS filter notation $K_s$
thoughout this paper for notational simplicity.}.

The distinction between red supergiants and yellow supergiants, for
our purpose at $(J-K) = 0.9$ mag, is useful here as it brings out the
nature of the contamination in the catalogs.  As one moves from yellow
to red supergiants, the contamination from Milky Way dwarfs and giants
decreases substantially.  \citet{neugent2012} found 22\% purity for
their yellow supergiant catalog and a 97\% purity for their red
supergiant catalog.  \citet{gonzalez2015}, performing a more detailed
spectral analysis, measured a 53\% purity for the red supergiants,
largely contaminated by carbon stars and MW giants.  At $M_K\lesssim
-9.5$ mag ($K\sim 9$ mag), the purity was $\gtrsim 95\%$, consistent
with \citet{neugent2012}.

The aforementioned studies did not cover the entire LMC:
\citet{neugent2012} covered $\sim$ 22 deg$^2$ of the LMC, about 60\% of
the relevant area, while \citet{gonzalez2015} covered a $\sim$ 3 deg$^2$
field at the densest part of the LMC.  In the region of overlap, the
latter analysis recovered about 3 times as many red supergiants as the
former analysis.  Both studies are also likely incomplete in regions
of very high stellar density (e.g., R136).  Reddening is not a factor
for the $J$ and $K$ bands, except for progenitors obscured by
molecular clouds. Otherwise, the highest extinction 3 arcmin$^2$ field
in the LMC has $E(B-V)\approx2.0$ mag, and only 0.26 deg$^2$ in the
200 deg$^2$ around the LMC has $E(B-V)\gtrsim 1$ mag; these correspond
to only 0.6 and 0.3 mag of extinction in the $K$-band, respectively.

\subsection{Constructing a LMC Red Supergiant Catalog}

We construct a catalog of luminous red supergiants in the LMC
following a similar analysis to that of \citet{gonzalez2015}.  We
begin with the 2MASS point source catalog within $3.5^\circ$ from
$\alpha,\delta = 79.5,-68.8$, and apply the following selection
criteria:

\begin{enumerate}
\item $K>9$ mag, $(J-K) > 0.9$ mag,

\item the pseudo-color cut of $0.1\ge q\ge 0.4$, where $q\equiv
  (J-H)-1.8 (H-K)$,

\item $10^{5} L_\odot < L < 10^{6} L_\odot$ ,

\item reject stars which have proper
  motions of $\sqrt{\mu_{ra}^2+\mu_{dec}^2} > 6$ mas yr$^{-1}$ with
  $\sqrt{\mu_{ra}^2+\mu_{dec}^2} > 3
  \sqrt{\sigma_{mu\_ra}^2+\sigma_{mu\_dec}^2}$ in the NOMAD catalog\citep{nomad2004}.

\end{enumerate}
The bolometric luminosity cut calculation follows \citet{neugent2012},
namely, the $(J-K)$ color is used to estimate the effective
temperature, and the effective temperature is in turn used to
calculate the bolometric correction.

This process yields 152 red supergiant candidates. 
This is smaller than the number of supergiants in either the catalogs
of \citet{neugent2012} or \citet{gonzalez2015} as these studies
go to much lower luminosities than we are concerned with here. This is
evident from Figure~2. The highest luminosity candidates are likely
all MW stars; the Neugent et al data show that 90\% of their
candidates at $K<7$ were MW stars. As we aim for completeness we find this acceptable.
In Figure~3 the candidate supergiants are shown overlaid on a stellar density map
of the LMC.

\begin{figure}
\hspace*{-0.5cm}\includegraphics[width=0.52\textwidth]{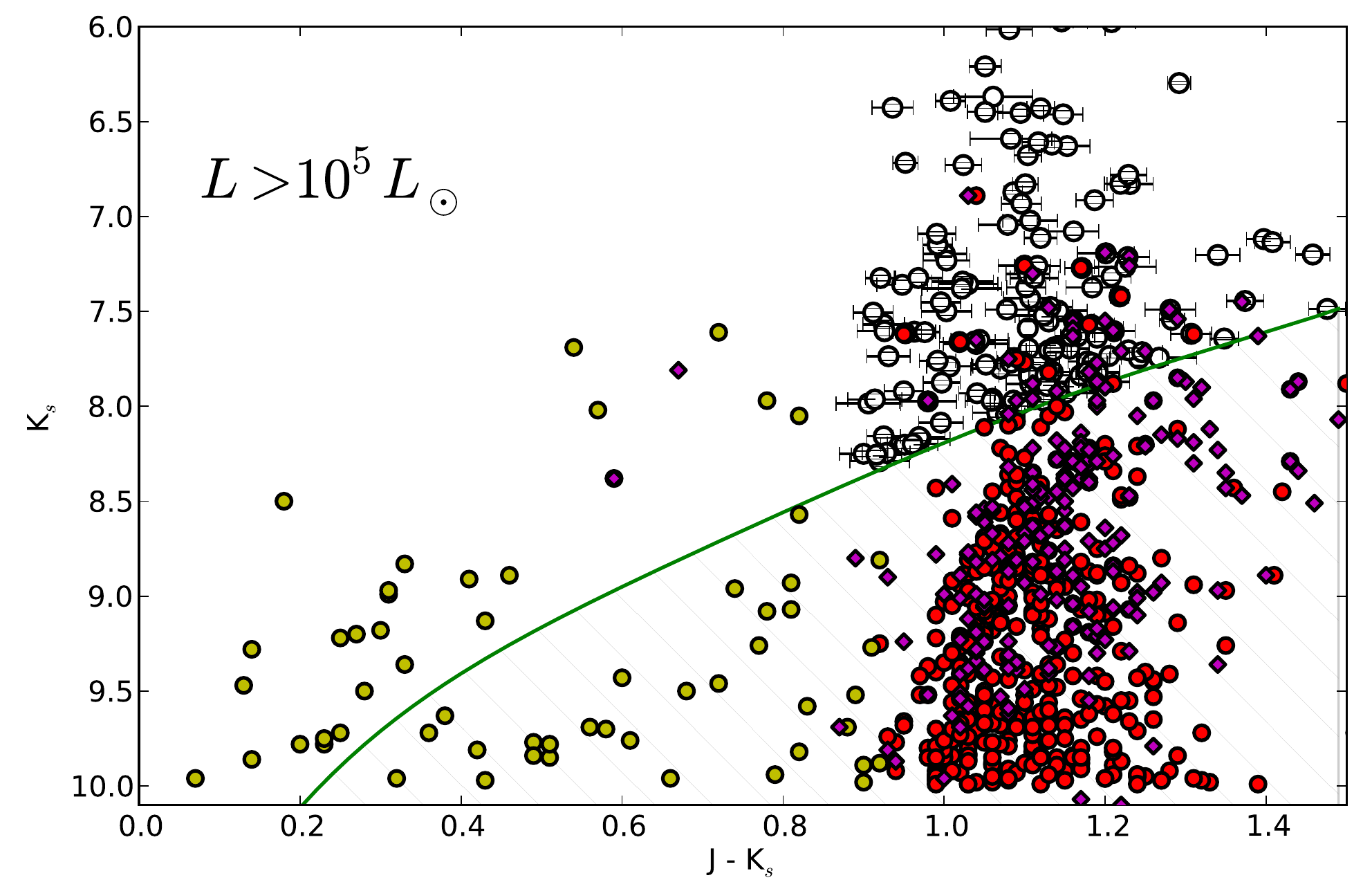}
\label{fig:supergiants}
\caption{ 2MASS $J-K$ vs.~$K$ diagram for the \citet{neugent2012}
  yellow supergiants (yellow circles) and red supergiants (red
  circles), \citet{gonzalez2015} red supergiants (purple diamonds),
  and the 152 supergiant candidates found here (white circles).  For
  our candidates, the uncertainties in both $(J-K)$ and $K$ are plotted;
  for $K$ they are smaller than the symbols.
  The line shows the dividing line for $10\textsuperscript{5}$ L$_\odot$.  }
\end{figure}

\section{Other failed supernova progenitors}

The red supergiant catalog has the advantage of being
well defined and motivated by observational evidence, 
but it does have uncertainties. These include
the calculation of the $10^5$ L$_\odot$ limit
and model uncertainties when  mapping the mass to luminosity.

There are more profound  uncertainties in the theory.
Smartt's analysis does not imply that only high luminosity red supergiants
could fail to explode.  The current theoretical models of core collapsing stars
either have islands of core-collapse to black holes at $\sim$20M$_\odot$ and $\sim$40M$_\odot$,
\citep{oconnor2011, pejcha2015}
or have most stars above  $\sim$20M$_\odot$ core collapsing to black holes 
\citep[with the interesting exception of an island of explosion at $\approx$26M$_\odot$]{sukhbold2015},
though examples of core collapse to black holes occur throughout the range 15M$_\odot$--120M$_\odot$
in the latter study.\footnote[2]{Throughout this paper, masses quoted are zero age main sequence masses.}
The lack of explosion depends on many parameters, notably metallicity 
\citep{pejcha2015} as the LMC averages half solar metallicity.
In theory, then, a direct collapse to black holes may occur in many observational
classes of massive stars:
yellow supergiants,  blue supergiants, 
luminous blue variable stars (LBVs), Wolf-Rayet (WR) stars, sgB[e], and 
more (see e.g.,  \citealt{kashiyama2015}). Fortunately, these 
classes of stars have been extensively studied in the LMC.

\section{The search for missing LMC supergiants in the DECam data}

The area covered in our DECam LMC campaign is shown in Figure~3.  The
DECam images were analyzed with the DES first cut reductions
\citep{sevilla2011, DESDM2012, desai2012, gruendl2016}, which include
producing astrometrically calibrated reduced images.
We visually inspected the locations of
the red supergiants in our catalog.  The supergiants were mostly saturated
in the images, so we could not investigate the brightening 
discussed in the previous section.  Our imaging and subsequent visual
inspection covered 144 supergiants, 95\% of the original catalog, and
all of these stars were recovered. We argue that this is the level
of confidence excluding a luminous red supergiant undergoing a failed SN 
in the LMC at the time of GW150914.

The catalogs of other possible failed SN progenitors are present in the literature.
We can check for the disappearence of less luminous
red supergiants and yellow supergiants 
using the catalog of \citet{neugent2012}: 813 of 846
(96\%) are in the imaged area and all of these are present in the
images. We can check for the disapperance of WR stars using the
catalog of \cite{hainich2014}, extensive but known not to be complete 
\citep{massey2015}: 105 of 108 (97\%) are in our imaged area and we can confirm
that 102 (97\%) are present.  The three that we cannot confirm are in
the very compact cluster R136, and are unresolved in our data.
We can check for the disappearence of LBVs using the stars from
\cite{smith2015}, which are all the confirmed, not highly reddened, LBVs in the LMC:
we recover 16 of 16 (100\%) in the DECam imaging.
We could have checked blue supergiants, including the interesting subclass sgB[e],
using the catalog in \cite{bonanos2009}.
As these catalogs are incomplete,
it is difficult to state how confident we are that these kinds of progenitors
did not undergo a failed SN in the LMV at the time of GW150914, but
given the uncertainty in theoretical predictions for which observational
classes of stars undergo failed SN, a reasonable compromise is to check the
known catalogs of potential progenitors.

\begin{figure}
\hspace*{-0.5cm}\includegraphics[width=0.52\textwidth]{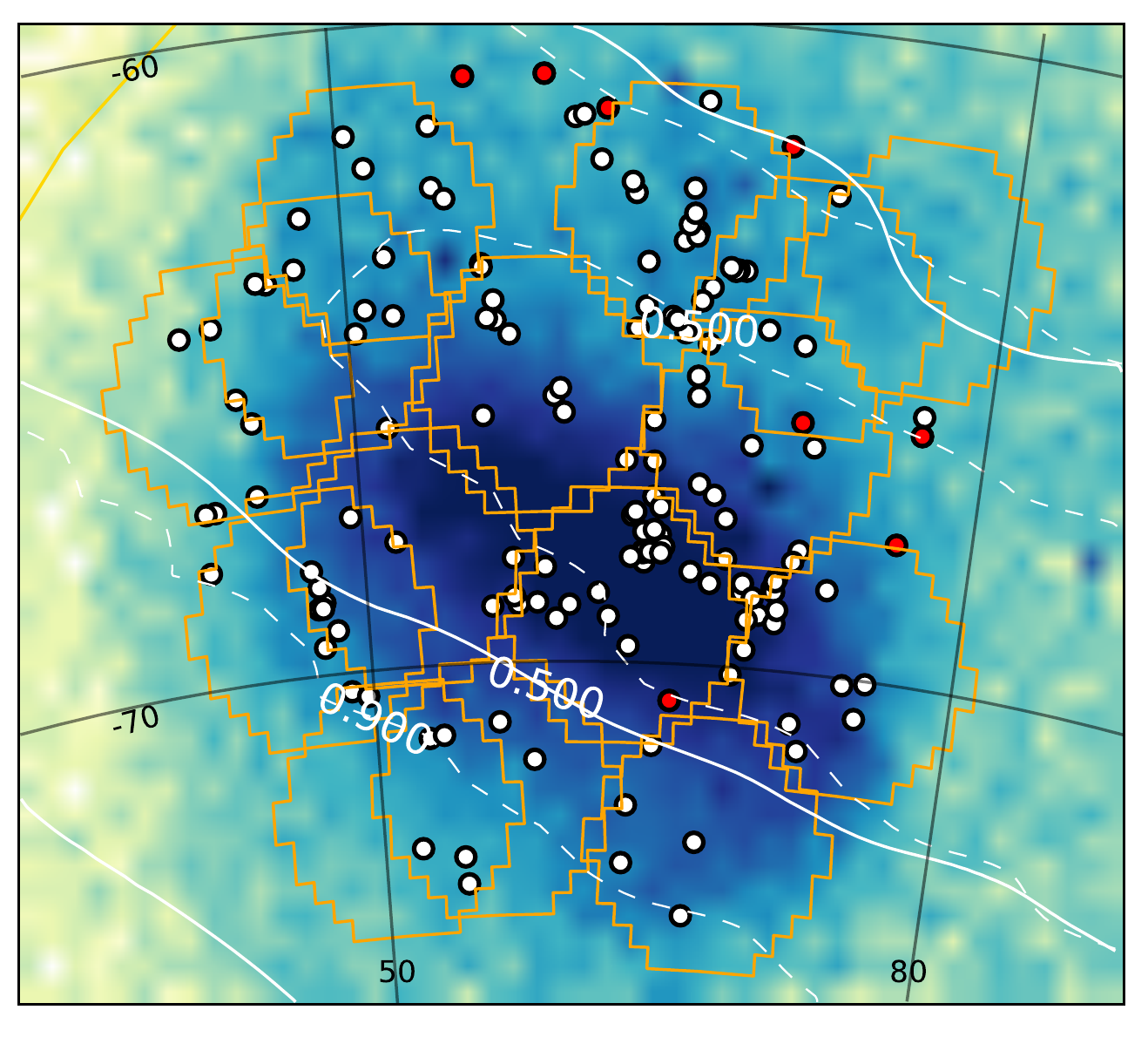}
\label{fig:lmcobs}
\caption{A map of the logarithm of 2MASS J-band star counts around the LMC
with the LIGO localization contours shown in white.
The DECam i-band images are shown as orange  camera outlines;
some of the z-band images are offset from these.
The white points are the luminous red supergiant catalog
developed in this paper, with those marked red not having
a visual inspection. Six are outside our imaging area. The
four remaining fell into chip gaps and/or on bad CCDs.}
\end{figure}

\section{Discussion and Conclusions}

GW150914 was first detected by a LIGO analysis sensitive to a burst of GW and 
the high probability localization contours enclosed the LMC.  Burst-like gravitational wave
signals could originate from the core-collapse of massive stars,
perhaps $\sim$20\% of which fail to explode as luminous SNe.  This motivated us
to search for a failed SN in the LMC.  We constructed
a catalog of 152 high luminosity LMC supergiants, of which 144 were
observed in our DECam imaging; all of these stars are still present
after the LIGO event.  It is unlikely that the then candidate event GW150914 originated
from a failed SN in the LMC.   The subsequent publication of the
GW150914 analysis shows that the GW event is consistent  
with a merging massive binary black hole model at $z\approx 0.09$
\citep{ligo2016}. 

The spatial uncertainty present in GW150914 will be a feature of all
non-electromagnetic core-collapse triggers.  Most models of a core
collapse, whether the final stage is a neutron star or a black hole,
include the formation of a neutrinosphere \citep[see][and references
therein]{scholberg2012}.  Thirty years ago the LMC core-collapse that
produced SN1987A was detected by two neutrino detectors, Kamiokande
and IMB \citep{kamiokande1987,imb1987}.  There are seven neutrino
detectors contributing to the SNEWS supernova early warning system
\citep{snews2011}, and the Super-Kamiokande neutrino detectors and the
IceCube neutrino telescope should detect an LMC core-collapse
unassisted \citep{ikeda2007, abbasi2011}.  Notably for this paper,
the MeV neutrino burst mode 
of IceCube did not trigger for 
$\pm 500$ seconds around the time of GW150914 \citep{iceligo2016}
which it would have for a core-collapse in the LMC.
The spatial localization of
the neutrino detectors is several degrees \citep{adams2013}---that
would be good enough to say the event likely occured in the LMC, 
but not where in the LMC it is located.  

The use of the luminous red supergiant catalog makes it possible to
perform a specific search without prior template imaging, and
therefore without difference imaging.  A sensible generalization of
this technique is to perform very shallow $g$ and $i$ band imaging
of very nearby galaxies to prepare template images for difference imaging;
$g$ band added to catch the very blue signature of a breakout shock.
Difference imaging in the crowded regions of the LMC will likely be
challenging, but would extend the discovery space to other possible
low luminosity core collapse progenitors, of which there are many.  
The durations between local group core collapses are measured in 
decades and we should be prepared to learn as much as possible when they do occur.

\acknowledgments

Funding for the DES Projects has been provided by the U.S. Department
of Energy, the U.S. National Science Foundation, the Ministry of
Science and Education of Spain, the Science and Technology Facilities
Council of the United Kingdom, the Higher Education Funding Council
for England, the National Center for Supercomputing Applications at
the University of Illinois at Urbana-Champaign, the Kavli Institute of
Cosmological Physics at the University of Chicago, the Center for
Cosmology and Astro-Particle Physics at the Ohio State University, the
Mitchell Institute for Fundamental Physics and Astronomy at Texas A\&M
University, Financiadora de Estudos e Projetos, Funda{\c c}{\~a}o
Carlos Chagas Filho de Amparo {\`a} Pesquisa do Estado do Rio de
Janeiro, Conselho Nacional de Desenvolvimento Cient{\'i}fico e
Tecnol{\'o}gico and the Minist{\'e}rio da Ci{\^e}ncia, Tecnologia e
Inova{\c c}{\~a}o, the Deutsche Forschungsgemeinschaft and the
Collaborating Institutions in the Dark Energy Survey.

The Collaborating Institutions are Argonne National Laboratory, the
University of California at Santa Cruz, the University of Cambridge,
Centro de Investigaciones Energ{\'e}ticas, Medioambientales y
Tecnol{\'o}gicas-Madrid, the University of Chicago, University College
London, the DES-Brazil Consortium, the University of Edinburgh, the
Eidgen{\"o}ssische Technische Hochschule (ETH) Z{\"u}rich, Fermi
National Accelerator Laboratory, the University of Illinois at
Urbana-Champaign, the Institut de Ci{\`e}ncies de l'Espai (IEEC/CSIC),
the Institut de F{\'i}sica d'Altes Energies, Lawrence Berkeley
National Laboratory, the Ludwig-Maximilians Universit{\"a}t
M{\"u}nchen and the associated Excellence Cluster Universe, the
University of Michigan, the National Optical Astronomy Observatory,
the University of Nottingham, The Ohio State University, the
University of Pennsylvania, the University of Portsmouth, SLAC
National Accelerator Laboratory, Stanford University, the University
of Sussex, and Texas A\&M University.

The DES data management system is supported by the National Science
Foundation under Grant Number AST-1138766.  The DES participants from
Spanish institutions are partially supported by MINECO under grants
AYA2012-39559, ESP2013-48274, FPA2013-47986, and Centro de Excelencia
Severo Ochoa SEV-2012-0234.  Research leading to these results has
received funding from the European Research Council under the European
Union’s Seventh Framework Programme (FP7/2007-2013) including ERC
grant agreements 240672, 291329, and 306478.

R.J.F.\ gratefully acknowledges support from NSF grant AST-1518052 and
the Alfred P.\ Sloan Foundation.
FS acknowledges financial support provided by São Paulo Research Foundation (FAPESP) under grants 2015/12338-1.

\end{document}